\documentclass[aps, pra, twocolumn, superscriptaddress, amsmath,  tightenlines, longbibliography]{revtex4-2}

\usepackage{dcolumn}
\usepackage{graphicx}
\usepackage{epstopdf}
\usepackage{mathrsfs}
\usepackage{subfigure}
\usepackage{booktabs}
\usepackage{array}
\usepackage{multirow}
\usepackage{amsmath,amsfonts}
\usepackage{physics}
\usepackage{dsfont}
\usepackage{amstext}
\usepackage{amssymb}
\usepackage{amsbsy}
\usepackage{bbm}
\usepackage{amsthm}
\usepackage{xcolor}
\usepackage[UTF8,scheme=plain]{ctex}
\usepackage[colorlinks,urlcolor=blue,linkcolor=blue,citecolor=blue]{hyperref}
\usepackage{soul}

\usepackage{url}
\usepackage[colorlinks]{hyperref}
\hypersetup{%
	plainpages=true,
	breaklinks=true,       %not default in dvips mode, so we must specify
	hypertexnames=false,  %not ideal, but needed when pagenums duplicate (`i' vs. `1')
	pageanchor=true,
	colorlinks=true,
	linkcolor={blue},
	citecolor={blue},
	urlcolor={blue},
	anchorcolor={black}
}

 \makeatletter

\newcommand{\Rmnum}[1]{\expandafter\@slowromancap\romannumeral #1@}
\makeatother

\begin{document}

\author{Hao-Yue Zhang}
\affiliation{School of Physics and Astronomy, Beijing Normal University, and Key Laboratory of Multiscale Spin Physics (Beijing Normal University), Ministry of Education, Beijing 100875, China}

\author{Rong-Hang Chen}
\affiliation{Beijing Computational Science Research Center, Beijing 100193, China}
\affiliation{School of Physics and Astronomy, Beijing Normal University, and Key Laboratory of Multiscale Spin Physics (Beijing Normal University), Ministry of Education, Beijing 100875, China}

\author{Qing Ai}
\email{aiqing@bnu.edu.cn}
\affiliation{School of Physics and Astronomy, Beijing Normal University, and Key Laboratory of Multiscale Spin Physics (Beijing Normal University), Ministry of Education, Beijing 100875, China}

\title{
Inferring Transition-Dipoles from {  Two-Dimensional Infrared Spectroscopy} with Deep Learning
}

%\abbreviations{2D IR,CNN,NHH}

\date{\today}

\begin{abstract}

%Two-dimensional infrared (2D IR) spectroscopy is a powerful nonlinear optical technique for probing molecular couplings, coherence, and dynamical processes. 
{  Two-dimensional infrared (2D IR) spectroscopy is a powerful nonlinear technique for probing vibrational couplings, anharmonicities, coherence, and molecular dynamics.}
Among the microscopic parameters encoded in {  2D IR spectra}, transition dipoles are fundamental quantities that characterize molecular structure and govern light-matter interactions. However, reconstructing these quantities from spectroscopic data remains a challenging inverse problem.
Here, we introduce a physics-informed convolutional neural network (CNN), Spec2DipoleNet, trained on {  2D IR spectra} generated via the non-Hermitian Hamiltonian (NHH) method. 
This multi-task CNN reconstructs the transition dipole matrices under physically-motivated masking constraints.
On the validation set, the model achieves a root-mean-square error (RMSE) of 0.085, an $R^2$ of 0.97, and a normalized cross-correlation of 0.98. It reliably recovers matrix elements associated with allowed transitions, while larger deviations are observed for nominally forbidden transitions.
{  Our results suggest that transition dipole operators can be quantitatively inferred from 2D IR spectra, highlighting the potential of physics-informed deep learning for extracting microscopic quantum parameters from multidimensional spectroscopic data.}

\end{abstract}

\maketitle

\section{Introduction}
\label{sec:introduction}
 
 The concept of two-dimensional spectroscopy was first established in nuclear magnetic resonance \cite{Aue-1976} and was later extended to ultrafast optical spectroscopy \cite{HYBL1998307,Asplund2000}.
 Two-dimensional coherent spectroscopy (2DCS) uses sequences of ultrafast pulses to generate nonlinear optical responses that are resolved along two frequency axes \cite{mukamel-2000}.
 The resulting two-dimensional spectra reveal couplings, coherences, and dynamical processes that are difficult to access using conventional spectroscopy \cite{WOS:000352259800029,WOS:000461530000001,Myers2008,WOS:000270766800043,Kearns2017,Oliver2015,Courtney2015,Song2019,Duchi2021,Nardin2013,WOS:000270766800043, WOS:000309566400010,WOS:000296494700045,WOS:000315525500002,WOS:000426497900006}. 
 Following early demonstrations in the electronic and infrared regimes \cite{HYBL1998307,Asplund2000}, 2DCS has developed into a broad family of techniques for investigating energy transfer, vibrational dynamics, and electronic relaxation in complex systems \cite{Engel2007,golonzka-2001,WOS:000269861400017}. 
 {  In the present work, we focus specifically on its infrared implementation, two-dimensional infrared (2D IR) spectroscopy.}
 
 Among the molecular parameters encoded in {  2D IR spectra}, transition dipole matrix elements are fundamental quantities for characterizing molecular structure and light-matter interactions \cite{herzberg2013molecular,grechko2012quantification,ahmed2006measurement,li2011direct}, and serve as key parameters in simulations of absorption spectra and excitation energy transfer processes \cite{muh2008crystal,raszewski2008light,king2001excited,muller2003ultrafast}. They are rarely measured directly, but are instead inferred from linear and nonlinear spectroscopic observables, such as absorption and emission intensities, radiative lifetimes, and coherent light-matter coupling signals \cite{grechko2012quantification,li2011direct,stancu2008measurement,bouchard2016branching,theisen2009femtosecond,Khalil2003JPCA,Demirdoven2002PRL,Golonzka2001}. Thus, 2D IR spectroscopy provides a route to quantitatively extract transition dipole strengths \cite{Demirdoven2002PRL,Khalil2003JPCA,grechko2012quantification}.

The interpretation of 2DCS relies on accurate theoretical modeling. A variety of frameworks have been developed, including response-function formalisms \cite{mukamel-1995,Deng2024PRA,WOS:000170647600002,Noblet2021}, hierarchical equations of motion \cite{tanimura-2020}, and direct numerical integration of the Schr\"odinger equation \cite{sardjan-2020,Eric2023}, among others \cite{Chen-2017,mukamel-2000,falvo-2009,cupellini-2020}.
 These methods offer different balances between physical accuracy, numerical cost, and range of applicability. 
 Among them, the non-Hermitian Hamiltonian (NHH) method offers an attractive compromise between computational efficiency and accuracy. Before being adapted to 2DCS simulations, the NHH method had been widely applied to {  parity-time symmetry} systems \cite{Bender1998PRL,Ashida2017,Feng2017}, photosynthetic light harvesting \cite{Olaya-Castro2008PRB,Celardo2012,Nesterov2013}, chemical reaction dynamics and avian magnetoreception \cite{JONES201090,Haberkorn1976,maeda_chemical_2008}, as well as negative refraction in M\"obius molecules \cite{Fang2016PRA}. When adapted to the simulation of 2DCS, this approach remains reliable in the short-population-time regime while significantly reducing computational complexity \cite{Khalil2003JPCA,Demirdoven2002PRL,Lemmer2015,Zhang2025}. This capability makes it particularly attractive for generating large-scale spectral datasets.

In parallel, the integration of deep learning with spectroscopy has attracted increasing attention. Machine learning has shown promising performance across a broad range of spectroscopic techniques, including ultraviolet-visible spectroscopy \cite{ye2019neural}, infrared spectroscopy \cite{yao2018tensormol,gastegger2017machine}, nuclear magnetic resonance \cite{paruzzo2018chemical}, X-ray absorption spectroscopy \cite{carbone2020machine,zheng2018automated}, and time-resolved spectroscopy \cite{kollenz2020unravelling}. 
However, most existing studies have primarily focused on classification tasks and coarse parameter inference, whereas the direct inference of physically meaningful quantities from the {  2D IR spectra} remains a challenging inverse problem that demands high predictive accuracy \cite{Parker2022,ren2022machine,wu2024unraveling}.

 Convolutional neural networks (CNNs) are a specialized class of deep neural networks for processing data with grid-like topology, such as images and time series \cite{Goodfellow2016Book,lecun2002gradient}. By exploiting local connectivity, weight sharing, and hierarchical feature extraction, 
 CNNs have achieved widespread success in domains ranging from computer vision and speech recognition to time-series analysis and the physical sciences 
  \cite{lecun2002gradient,Krizhevsky2012,He2016,Huang2017,LeCun2015Nature,Goodfellow2016Book,Alzubaidi2021Review,Cong2023Review,hinton2012deep,kim2014convolutional,bai2018empirical,carleo2017solving}. CNNs are particularly attractive for spectroscopic applications due to the local inductive bias, parameter efficiency, and robustness in limited-data regimes \cite{Cong2023Review}.

In this work, we address the inverse problem of recovering the transition dipole moment operator from {  2D IR spectra}. To this end, we construct a systematically generated dataset of simulated {  2D IR spectra} using the NHH method and develop a physics-informed multi-task neural network, Spec2DipoleNet, to infer the transition dipole operator directly from spectral data. 
The model jointly predicts the energy-level structure and the transition dipole matrix, and employs physically motivated masking constraints to improve reconstruction accuracy and interpretability.
Furthermore, we decouple structure-prediction errors from regression errors, allowing a rigorous evaluation of the intrinsic regression performance and identifiability of transition dipole recovery from {  2D IR spectra}.

This paper is organized as follows. In the next section, we introduce the NHH approach used to generate the {  2D IR spectra} for training and validating the deep learning model. In Sec.~\ref{sec:Spec2DipoleNet}, we describe  Spec2DipoleNet, an architecture for inferring transition dipoles from {  2D IR spectra}. In Sec.~\ref{sec:Training}, we present the training results and evaluate its performance on the validation set. Finally, we summarize our main findings and discuss the scope of the present framework in Sec.~\ref{sec:Conclusion}.

\section{Construction of a Two-Dimensional Infrared Spectroscopy Dataset Using the Non-Hermitian Hamiltonian Method} \label{sec:Dataset Preparation}

\begin{figure}[h]
            \includegraphics[width=8.4 cm]{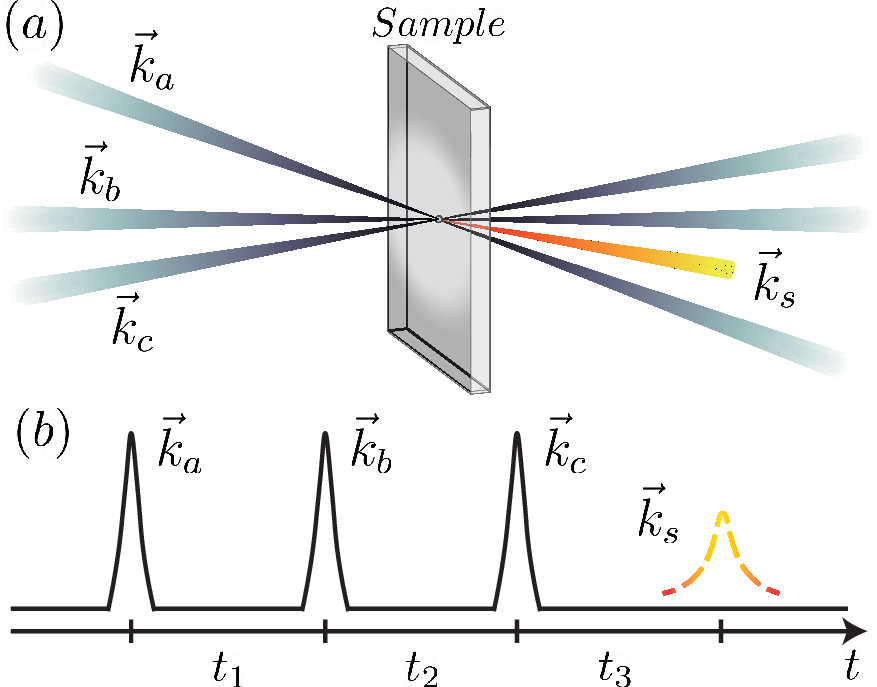}
            \caption{Schematic diagram of {  2D IR spectroscopy}. (a) Spatial configuration of the {   first two pump pulses and the third probe pulse} incident on the sample, together with the emitted signal. (b) Temporal sequence of pulses.}
            \label{fig:Two-Dimensional Spectroscopy}
\end{figure}

{  2D IR spectroscopy} uses a series of sequenced pulses incident on a sample \cite{Jonas2003}. The system undergoes coherent evolution during the interpulse interval and emits signals in a specific phase-matching direction.  The signals are mapped to the frequency domain by double Fourier transformation with respect to the first and last intervals, yielding a {  2D IR spectra}. This technique is schematically illustrated in Fig.~\ref{fig:Two-Dimensional Spectroscopy}, where the blue cones represent {  two pump pulses and one probe pulse},
while the orange cone indicates the emitted signal. $\vec{k}_\alpha$~($\alpha=a,b,c$) and $\vec{k}_s$ denote the wave vectors of pulses and the emitted signal, respectively, while $t_i~(i=1,2,3)$ denote the delays between successive pulses, whereas $t_3$ is the detection time following the third pulse.

The NHH method makes a trade-off between computational complexity and accuracy,  providing reliable results in the short-population-time regime \cite{Khalil2003JPCA,Demirdoven2002PRL,Lemmer2015,Zhang2025}.
By dropping the quantum-jump terms in the quantum master equation, the NHH in the interaction picture is written as
\begin{align}
    H^{\rm I}_{\rm NH}= & H^{\rm I}_{\rm S}-\frac{i}{2} \sum_{\beta \neq g} ( \Gamma_{1}^{\beta}A_{g\beta}^{\dagger}A_{g\beta}+\Gamma_{2}^{\beta}A_{\beta g}^{\dagger}A_{\beta g}  \nonumber\\ 
    &+\gamma_{\beta}A_{\beta\beta}^{\dagger}A_{\beta\beta}),
\end{align}
where  $A_{\alpha\beta}=\vert \alpha\rangle\langle \beta\vert$ denotes the quantum-jump operator, $\vert g \rangle$ is the ground state, $\Gamma_{1}^{\beta}$ and $\Gamma_{2}^{\beta}$ are the downhill and uphill relaxation rates, respectively, $\gamma_{\beta}$s are the pure-dephasing rates, and $H^{\rm I}_{\rm S}$ denotes the system Hamiltonian in the interaction picture. Furthermore, during {  the pump and probe pulses,} the Hamiltonian is treated within the square-pulse approximation, while relaxation and dephasing are neglected. Therefore, the Hamiltonian is written as
\begin{align}
    H^{\rm I}_{p}=-\frac{1}{2}
    \sum_{\alpha \neq \beta} \Omega_{\alpha \beta} e^{-i\vec{k}_{p}\cdot\vec{r}}\vert \alpha \rangle\langle \beta \vert
    +{\rm h.c.},
\end{align}
where the Rabi frequencies are $\Omega_{ij}=-\vec{d}_{ij}\cdot\vec{E}$, with $\vec{E}$ being the electric field of the laser pulse, and $\vec{d}_{ij}$ being the transition dipole moment between states $i$ and $j$.
 
 Denoting the interaction-picture evolution operators during the {  optical pulses} and during field-free evolution by $U^{\rm I}_p$ and
 $U^{\rm I}_f$, respectively, the final state of the system is written as \cite{Zhang2025, Cai2022PRL}
    \begin{align}
        &\vert\psi^{\rm I}(t_{3},t_{2},t_{1})\rangle=\nonumber\\
&U^{\rm I}_{f}(t_{1}\!+\!t_{2}\!+\!t_{3}+\delta t_{a}\!+\!\delta t_{b}\!+\!\delta t_{c},t_{1}\!+\!t_{2}\!+\!\delta t_{a}\!+\!\delta t_{b}\!+\!\delta t_{c})\nonumber\\
        &{\times}U^{\rm I}_p(t_{1}+t_{2}+\delta t_{a}+\delta t_{b}+\delta t_{c},t_{1}+t_{2}+\delta t_{a}+\delta t_{b})\nonumber\\
        &{\times}U^{\rm I}_{f}(t_{1}+t_{2}+\delta t_{a}+\delta t_{b},t_{1}+\delta t_{a}+\delta t_{b})   \nonumber\\   
        &{\times}U^{\rm I}_p(t_{1}+\delta t_{a}+\delta t_{b},t_{1}+\delta t_{a})U^{\rm I}_{f}(t_{1}+\delta t_{a},\delta t_{a})\nonumber\\   
        &{\times}U^{\rm I}_p(\delta t_{a},0)\vert\psi(0)\rangle,
    \end{align}
    where, for $U_\alpha^{\rm I}(t_f,t_i)$ ($\alpha=f,p$), $t_i$ and $t_f$ denote the beginning and end of the corresponding evolution interval, and $\delta t_p$ ($p=a,b,c$) {  denotes the duration of pulse $p$.}
    The density matrix in the interaction picture after evolution over the time interval $t_3$ is given by $\rho^{\rm I}(t)=\vert\psi^{\rm I}\left(t_{3},t_{2},t_{1}\right)\rangle \langle\psi^{\rm I}\left(t_{3},t_{2},t_{1}\right)\vert$. The corresponding polarization is then expressed as $P(t)={\rm Tr}(\mu^{\rm I}(t) \rho^{\rm I}(t))$.

{  To construct a large dataset for inferring transition dipole moments from 2D IR spectra,} the NHH method is employed to compute the spectral response. 
Due to the phase-matching condition, the third-order polarization in the rephasing (RP) direction is emitted along
$\vec{k}_{s}=-\vec{k}_{a}+\vec{k}_{b}+\vec{k}_{c}$
and is denoted as $\vec{P}_{\rm RP}(t_{3}, t_{2}, t_{1})$. 
By performing double Fourier transforms with respect to $t_1$ and $t_3$, the RP signal is given by
\begin{align}
    \vec{P}_{\rm RP}(\omega_{3}, t_{2}, \omega_{1})
    = \mathcal{F}_{t_3 \to \omega_3} \mathcal{F}_{t_1 \to \omega_1}
    \left[\vec{P}_{\rm RP}(t_{3}, t_{2}, t_{1})\right].
\end{align}
Similarly, the third-order polarization in the nonrephasing (NR) direction, denoted as $\vec{P}_{\rm NR}(t_{3}, t_{2}, t_{1})$, is emitted along
$\vec{k}_{s} = \vec{k}_{a} - \vec{k}_{b} + \vec{k}_{c}$
or
$\vec{k}_{s} = \vec{k}_{a} + \vec{k}_{b} - \vec{k}_{c}$,
depending on the specific pulse ordering and phase-matching configuration.
The primary advantage of this approach lies in its relatively low computational cost, making it well suited for large-scale data generation. However, it is most accurate in the short-population-time regime, while differences may appear when describing long-population-time dynamics or steady-state behavior.
%In principle, the dataset can be extended to include spectra at multiple $t_2$ values. However, this would require either a set of 2DS evaluated at different $t_2$ values or a representation in which $t_2$ is explicitly included as an additional input variable, since $t_2$ constitutes an independent degree of freedom. 

{ 
The restriction to $t_2=0$ is mainly motivated by the computational cost
and the scope of the present study. Unlike the $t_1$ and $t_3$ intervals,
which are scanned and Fourier transformed to construct a single 2D IR
spectrum, sampling the population time $t_2$ produces a series of 2D IR
spectra, $P(\omega_3,t_2,\omega_1)$. A quantitative characterization of
$t_2$-dependent dynamics, such as relaxation rates or coherent-oscillation
frequencies, therefore requires spectra evaluated at multiple $t_2$ values,
which would substantially increase the size of the training dataset and the
dimensionality of the model input. Since the present work focuses on
reconstructing the transition dipole operator, which is a static molecular
quantity, we restrict the study to $t_2=0$. In principle, the framework can
be extended either by using 2D IR spectra evaluated at multiple $t_2$ values
as input or by explicitly including $t_2$ as an additional input variable.

Fixing $t_2=0$ excludes the dynamical information accumulated during the
population period, including coherent evolution, population relaxation,
energy transfer, spectral diffusion, and chemical exchange. The intrinsic
$t_2$-dependent modulation of the spectral signal should, however, be
distinguished from the systematic effect introduced by the NHH
approximation. Since the quantum-jump terms are neglected, the NHH treatment
introduces an additional attenuation of the absolute signal with increasing
$t_2$. As demonstrated in our previous comparison with the response-function
formalism~\cite{Zhang2025},
this attenuation mainly affects the absolute signal scale, whereas the
relative information among spectral peaks is predominantly preserved within the same set of Liouville pathways at long
population times. The corresponding systematic limitation of the NHH
approximation therefore lies primarily in the population normalization and
long-time steady-state behavior.
}

The dataset is generated in MATLAB. 
It comprises 11 sub-datasets of small- and medium-sized systems containing $N=3-10$ energy levels.
The main parameters and sample sizes of these sub-datasets are summarized in Tab.~\ref{tab:SamAndLevel}.
Each system consists of a ground state, singly excited states, and doubly excited states, where the number of doubly excited states may be zero. The number of singly excited states is set to $n_e = 2$ or $3$, while the doubly excited states are combinations of singly excited states, with $n_f \in [0,~6]$. For a given $n_e$, the maximum number of doubly excited states is $n_e (n_e + 1)/2$. 
Values of $n_f$ below this maximum correspond either to degeneracy in the doubly excited manifold or to cases in which only a subset of these states is accessible because of the finite pulse bandwidth.
Accordingly, the pulse bandwidth is chosen to be sufficiently broad to cover all singly excited states, while remaining insufficient to induce direct transitions from the ground state to the doubly excited manifold.

\begin{table}[h] 
\centering
           %\fontsize{12}{14}\selectfont
           %\resizebox{8.45cm}{!}{%
\begin{tabular}{c c c c}

\toprule
           %\hline
\quad\quad\quad$n_t$\quad\quad\quad & \quad\quad\quad$n_e$\quad\quad\quad & \quad\quad\quad$n_f$\quad\quad\quad & \quad\quad\quad$N_s$\quad\quad\quad   \\ 
\midrule
           %\hline
           10 & 3 & 6 & 1000 \\ 
           9  & 3 & 5 & 1000 \\ 
           8  & 3 & 4 & 1000 \\ 
           7  & 3 & 3 & 1000 \\ 
           6  & 2 & 3 & 500  \\ 
           6  & 3 & 2 & 500  \\ 
           5  & 2 & 2 & 500  \\ 
           5  & 3 & 1 & 500  \\ 
           4  & 2 & 1 & 500  \\ 
           4  & 3 & 0 & 500  \\ 
           3  & 2 & 0 & 1000 \\ 
\bottomrule
           %\hline
\end{tabular}%
           %}
\caption{The total number of states $n_t=n_e+n_f+1$, the number of singly excited states $n_e$, the number of doubly excited states $n_f$, and the sample size for different classes of systems $N_s$.}
\label{tab:SamAndLevel}
\end{table}

The lowest singly excited state is sampled within the frequency range of $[300, 2300]~\mathrm{cm^{-1}}$ \cite{Parker2022,Renger2000}. 
%, thereby ensuring that spectral features remain resolvable. Excessive broadening leads to peak overlap and loss of spectral resolution.
{ 
For the singly excited states, the relaxation and pure-dephasing rates are
independently sampled for each energy level, with their sum constrained to
remain below half of the minimum energy-level spacing. For the
doubly excited states, the corresponding relaxation and pure-dephasing rates
are generated based on those of the associated singly excited states, with
an additional random fluctuation of $\pm 10\%$. The resulting transitions
therefore exhibit state-dependent homogeneous broadening. The imposed constraint excludes the regime of excessive broadening in which
neighboring spectral features become largely unresolvable. Accordingly, the
present dataset represents relatively sparse vibrational manifolds with
resolved or moderately overlapping spectral features.
}

The transition dipole elements $\vec{\mu}_{m,mn}$ correspond to the allowed transitions $\vert m \rangle \leftrightarrow \vert mn \rangle$. 
The elements $\vec{\mu}_{m,nn}$ often vanish or are neglected under standard assumptions. However, experimental evidence suggests that these transitions should instead be treated as weakly allowed rather than strictly forbidden transition \cite{Khalil2003JPCA,Demirdoven2002PRL,Golonzka2001}.
When the singly  and doubly excited states are treated as product states, the relations $\vec{\mu}_{m,mm} = \sqrt{2}\,\vec{\mu}_{0,m}$ and $\vec{\mu}_{m,mn} = \vec{\mu}_{0,n}$ hold.
 In more realistic scenarios, these relations are only approximate. Accordingly, the transition dipole matrix elements between singly and doubly excited states are generated from their relations to the ground-to-singly excited transitions. Additional random fluctuations of $\pm 10\%$ are introduced to mimic deviations from the ideal relations arising from internal coupling and excitonic mixing.
{  The generated transition dipole amplitudes are expressed in dimensionless relative units. Before training, each transition dipole matrix is normalized separately using the median of the absolute values of its nonzero matrix elements. }

\begin{figure}[h]
            \includegraphics[width=8.4cm]{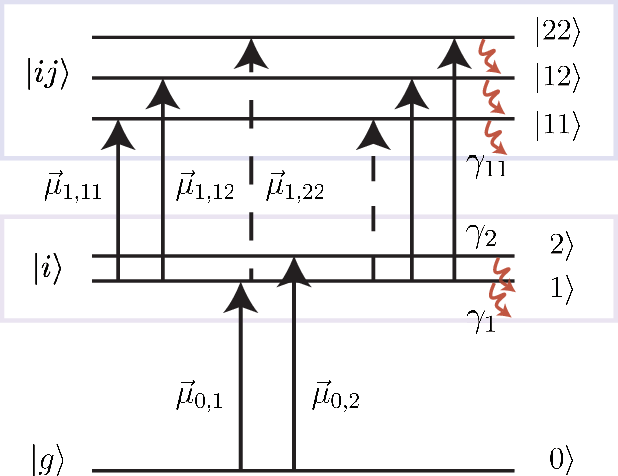}
            \centering
            \caption{Schematic structure of the six-level system. The solid arrows denote the allowed transitions, the dashed arrows denote the forbidden transitions, and the red arrows indicate the decoherence processes.}
            \label{fig:Molecule System}
\end{figure}

The dataset includes a class of six-level systems with a level structure similar to that of the Rh(CO)$_2$C$_5$H$_7$O$_2$ molecule \cite{Khalil2003JPCA,Demirdoven2002PRL,Golonzka2001}, {  providing a physically motivated reference for part of the parameter space, while the energy levels and dissipation rates are varied across individual samples}. In this case, $n_e = 2$, and the number of doubly excited states reaches its maximum value, i.e., $n_f = 3$, as shown in Fig.~\ref{fig:Molecule System}. The transition dipole moment is denoted by $\vec{\mu}_{i,f}$ for the transition $\vert i \rangle \leftrightarrow \vert f \rangle$, where $\vert i \rangle$ and $\vert f \rangle$ may belong to either the singly  or doubly excited manifolds.

Under the dipole and square-pulse approximation, the interaction Hamiltonian during the laser pulse is given by
\begin{equation}
{V}=-\vec{\mu}\cdot\vec{E}.
\end{equation}
In the laboratory frame, for a pulse propagating along the $X$ direction, the electric field $\vec{E}(t)$ lies in the $YZ$ plane. Each light--matter interaction is determined by the component of the transition dipole moment along the electric-field polarization.

Experimentally, {  2D IR spectra} with different polarization configurations can be obtained by controlling the field polarizations, examples include, $\vec{P}_{ZZYY}$, $\vec{P}_{ZYZY}$, $\vec{P}_{ZYYZ}$, and $\vec{P}_{ZZZZ}$. The subscripts $Z$ and $Y$ denote linear polarization along the corresponding axes, and the interaction sequence is read from right to left. For linearly-polarized fields along the $Z$ or $Y$ direction, the interaction reduces to
\begin{align}
V_Z &= -\vec{\mu} \cdot \vec{E}_Z = -\mu_Z E_Z, \\
V_Y &= -\vec{\mu} \cdot \vec{E}_Y = -\mu_Y E_Y.
\end{align}
In addition, these polarization configurations are not independent, but satisfy
\begin{equation}
\vec{P}_{{ZZZZ}} = \vec{P}_{{ZZYY}} + \vec{P}_{{ZYZY}} + \vec{P}_{{ZYYZ}}.
\end{equation}
% A unique solution of the full 3D orientation of a selected electronic e within the molecular frame requires measurement of the angles between at least three linearly independent vibrational tdm vectors and a single selected electronic tdm e (Femtosecond polarization resolved spectroscopy: A tool for determination of the threedimensional orientation of electronic transition dipole moments and identification of configurational isomers)
Although the $Z$ and $Y$ axes are equivalent under rotational symmetry, different polarization sequences correspond to distinct Liouville pathways. In particular, $\vec{P}_{{ZYZY}}$ and $\vec{P}_{{YZYZ}}$ are not equivalent.

{ 
In the generation of the dataset, orientational averaging is not included. Such averaging accounts for the random distribution of molecular orientations in isotropic ensembles. 
However, orientational averaging causes the spectral amplitudes to depend not only on the transition dipole strengths but also on their projections onto the pulse polarizations across a large ensemble of randomly oriented molecules. This weakens the correspondence between the spectra and the transition dipole matrix.
To avoid this additional ambiguity in the present inverse problem, we neglect orientational disorder and assume a fixed molecular orientation.}

{ 
For the $\vec{P}_{ZZZZ}$ polarization configuration considered here, the
spectral response depends on the projections of the transition dipoles along
the laboratory $Z$ axis. Accordingly, the target transition dipole matrix
can be interpreted as the corresponding $Z$-projected transition dipole
amplitudes. In principle, analogous $\vec{P}_{XXXX}$ and
$\vec{P}_{YYYY}$ measurements can provide the corresponding $X$- and
$Y$-projected amplitudes, from which the transition dipole strengths and
their projection angles relative to the laboratory axes can be obtained,
up to the sign ambiguity discussed below.

The signs of the Cartesian projections cannot in general be uniquely
determined from these measurements, because for both the all-parallel and
mixed polarization configurations considered here, each transition dipole
projection enters the relevant third-order pathways in pairs. Consequently,
reversing the sign of a projected component does not change the corresponding
spectral amplitude, and the signs of the Cartesian projections cannot be
distinguished from these measurements.
}
Without loss of generality, the polarization direction considered in the
present dataset is chosen as the $Z$ axis.

\begin{figure}[h]
                \includegraphics[width=8.45cm]{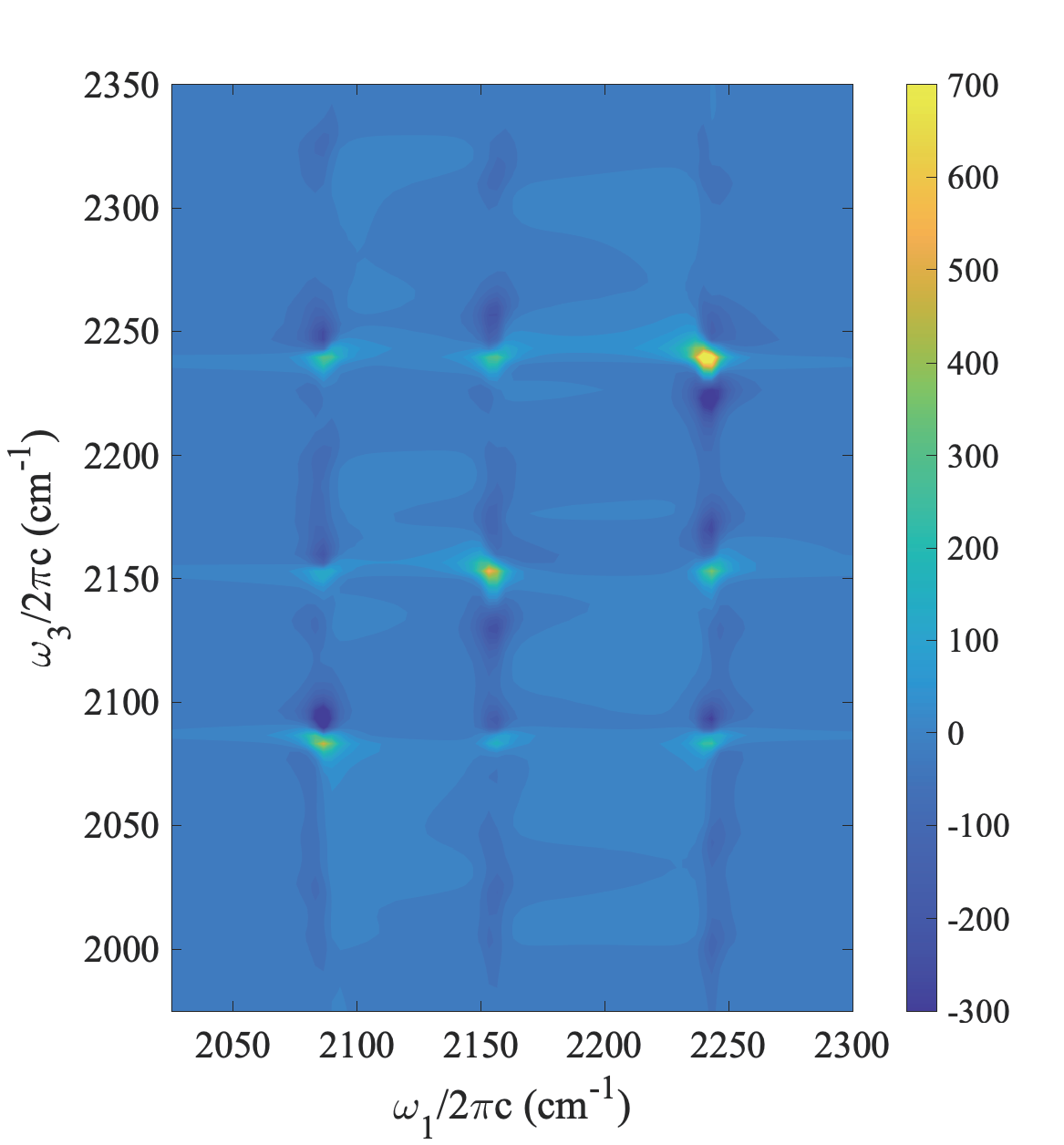}
                \centering
                \caption{Simulated $\vec{P}_{{ZZZZ}}$ {  2D IR spectrum} of a ten-level system.}
                \label{fig:ZZZZ2DS}
\end{figure}

Figure~\ref{fig:ZZZZ2DS} shows a simulated {  2D IR spectrum} of a ten-level system for the $\vec{P}_{{ZZZZ}}$ polarization configuration. The system consists of one ground state, three singly excited states, and six doubly excited states. The singly excited energies are $\omega_{1,0}=2086~\mathrm{cm}^{-1}$, $\omega_{2,0}=2155~\mathrm{cm}^{-1}$, and $\omega_{3,0}=2242~\mathrm{cm}^{-1}$, while the doubly excited energies are $\omega_{11,0}=4181~\mathrm{cm}^{-1}$, $\omega_{12,0}=4246~\mathrm{cm}^{-1}$, $\omega_{13,0}=4337~\mathrm{cm}^{-1}$, $\omega_{22,0}=4288~\mathrm{cm}^{-1}$, $\omega_{23,0}=4415~\mathrm{cm}^{-1}$, and $\omega_{33,0}=4469~\mathrm{cm}^{-1}$.
Along the diagonal, three positive peaks appear at $(2086,\,2083)~\mathrm{cm}^{-1}$, $(2153,\,2153)~\mathrm{cm}^{-1}$, and $(2240,\,2240)~\mathrm{cm}^{-1}$, corresponding to the transitions from the ground state to the singly excited states. Off-diagonal positive peaks are also located near these energies, reflecting distinct coherence frequencies during the evolution periods of $t_1$ and $t_3$.
Negative peaks originate from 18 possible downhill transitions between the doubly excited states and the singly excited states. Several of these transitions have nearly identical frequencies. For example, $\vert 11 \rangle {\rightarrow} \vert 1 \rangle$ and $\vert 13 \rangle {\rightarrow} \vert 3 \rangle$ are degenerate, while $\vert 12 \rangle {\rightarrow} \vert 3 \rangle$ and $\vert 22 \rangle {\rightarrow} \vert 1 \rangle$ differ by only $2~\textrm{cm}^{-1}$. Consequently, the corresponding negative peaks overlap in the {  2D IR spectrum} and therefore can not be fully resolved.

After the interaction with the first {  pump pulse}, the system is prepared in a coherent superposition of the ground state and the singly excited states. During the period of $t_1$, this coherence evolves at the transition frequencies between the singly excited states and the ground state. 
After the interaction with the {  probe pulse}, the system is driven into superpositions of the ground and singly excited states, or of the singly excited and doubly excited states. The emitted signal therefore contains contributions at the transition frequencies between the singly excited states and the doubly excited states. 
Consequently, the horizontal axis $\omega_{1}$ reflects the energy differences between the singly excited states and the ground state, whereas, for the negative peaks, the vertical frequency $\omega_{3}$ reflects energy differences between the doubly  and singly excited states.

{ 
For isotropic ensembles, orientational averaging is performed over all
molecular orientations, so that the absolute laboratory-frame Cartesian
projections of individual transition dipoles are no longer directly
retained.
One possible extension of the present framework to this regime would be to
represent the transition dipoles as molecular-frame vectors in the forward
model and perform orientational averaging during dataset generation.
The corresponding inverse problem would then need to target rotationally
invariant quantities, such as transition-dipole strengths and pairwise scalar
products, rather than laboratory-frame projections.
Different relative polarization sequences weight different combinations of these quantities and may therefore provide complementary constraints.
However, whether the underlying rotationally invariant quantities can be
uniquely recovered remains to be established, since different Liouville
pathways contribute with distinct orientational and dynamical weighting.
}

    \section{Spec2DipoleNet}
    \label{sec:Spec2DipoleNet}
    
    The {  2D IR spectra} are represented as two-dimensional matrices that depend only on $\omega_1$ and $\omega_3$. Accordingly, each {  2D IR spectrum} is treated as a single-channel input to the CNN.
    
\begin{figure}[h]
        \includegraphics[width=8.45cm]{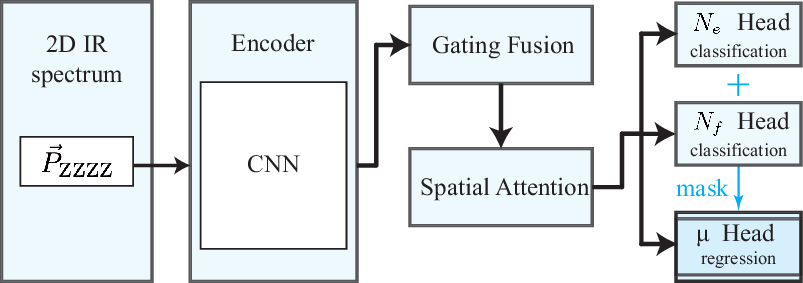}
        \centering
        \caption{Schematic architecture of Spec2DipoleNet, consisting of four main components.}
        \label{fig:CNN Model}
    \end{figure}

    Figure~\ref{fig:CNN Model} illustrates the architecture of the neural network used to predict the transition dipole matrix. The model comprises four main components. A convolutional backbone is first employed to encode the input {  2D IR spectra} into high-level feature representations. The extracted features are subsequently refined through a gated feature-fusion module and a spatial attention module. Finally, three prediction heads are used to produce the outputs, i.e., the $N_e$ head for the classification of the number of singly excited states, the $N_f$ head for the presence of the doubly excited states, and the $\mu$ head for the matrix of the transition dipole moment.
    
    The input to Spec2DipoleNet is a $991 \times 991$ $\vec{P}_{ZZZZ}$ {  2D IR spectrum} in the frequency range $[0,\,3300]\,\mathrm{cm}^{-1}$. 
    {  Before being fed into the network, each spectrum is normalized by its maximum absolute amplitude and subsequently multiplied by $10^{3}$ for numerical scaling.}
    The model adopts the Swish activation function with $\beta = 0.5$, which enhances nonlinear representation capability while preserving negative-valued features in the spectrum.
    
    To improve generalization and reduce overfitting, stochastic data augmentation is applied to the input {  2D IR spectra} during training with probability $p=0.8$. Specifically, each sample is randomly rescaled by a factor drawn from $[0.9,\,1.1]$, shifted by up to $\pm 6$ pixels along each frequency axis, and subjected to a cutout operation in which a rectangular region covering $8\%$ of the spatial extent is replaced by the sample mean. These random translations mimic global shifts in the transition frequencies and thereby broaden the sampled energy distribution. These augmentations preserve the global spectral profile while improving robustness to amplitude fluctuations, small peak shifts, and local perturbations.
    
    The encoder extracts spectral features using a CNN with residual connections. It consists of a shallow feature-extraction module (PreConvolution), followed by a stack of residual convolutional blocks (ResidualBlocks).
    These residual blocks progressively extract higher-level representations. The encoder is organized into four stages. The first stage contains a single residual block, while each of the remaining stages contains three residual blocks. Each  block comprises two $3 \times 3$ convolutional layers, with GroupNorm and Swish activation.
    
    Across stages, the number of channels increases while the spatial resolution decreases. 
    The detailed architectural parameters are summarized in Tab.~\ref{tab:Spec2DipoleNet}, where the design balances the computational cost and model capacity given the available GPU resources and the input size~\cite{tan2019efficientnet}.
    
    \begin{table}[h]
        
        \centering
        \resizebox{8.45cm}{!}{%
        \begin{tabular}{c c c c c}
        \toprule
        \textbf{Stage} & \textbf{Module} & \textbf{Channels} & \textbf{Kernel} & \textbf{Stride} \\ 
        \midrule
        Input & Polarization branches (\( \times 3 \)) & 1 & \textendash & \textendash \\ 
        PreConv & Conv + GN + Swish & 16 & 3x3 & 1 \\ 
        Stage1 & Residual Block \(\times 1\) & 16 & 3x3 & 1 \\ 
        Stage2 & Residual Block \(\times 3\) & 32 & 3x3 & 2, 1 \\ 
        Stage3 & Residual Block \(\times 3\) & 64 & 3x3 & 2, 1 \\ 
        Stage4 & Residual Block \(\times 3\) & 128 & 3x3 & 2, 1 \\ 
        Fusion & Gated Fusion & 128 & 1x1 & \textendash \\ 
        Pooling & Peak Attention Pooling & 128 & \textendash & \textendash \\ 
        Heads & Multi-task prediction heads & \textendash & \textendash & \textendash \\ 
        \bottomrule
        \end{tabular}%
        }
        
        \caption{Network architecture and selected parameters.}
        \label{tab:Spec2DipoleNet}
    \end{table}
    
    After extracting deep features, a spatially varying softmax-based gating mechanism is applied to adaptively fuse channel-wise features. The fused feature map is added to the encoder output as a residual enhancement. 
    The resulting features are fed into an attention pooling module, which produces both a global discriminative representation for classification tasks and a high-dimensional representation for the regression task.
    
    At the output stage, Spec2DipoleNet adopts a multi-task learning framework with three prediction heads. The $N_e$ head outputs logits corresponding to the number of the singly excited states. The $N_f$ head predicts the probability that each possible doubly excited state is present. The $\mu$ head regresses the matrix of the transition dipole moment.
    
    To enforce physical constraints, a mask matrix is introduced in the regression head. Specifically, only the off-diagonal elements in the upper triangular part of the symmetric dipole matrix are predicted. Meanwhile, transitions from the ground state to the doubly excited states, the transitions between two singly excited states, and the transitions between two doubly excited states are excluded. This mask incorporates prior physical knowledge, reducing parameter redundancy and improving the physical consistency of the predictions.
    The construction of the mask depends on the predicted or true values of $N_e$ and $N_f$. The final dipole matrix is obtained by symmetrizing the predicted matrix with its Hermitian conjugate.
    
    The model is trained using the AdamW optimizer with a batch size of 32 for a total of 160 epochs. The learning rate is linearly warmed up over the first 10 epochs, followed by a cosine annealing schedule with warm restarts. The initial restart period is set to 50 epochs and doubles after each restart. The 160th epoch corresponds to the end of the second restart cycle after the warm-up period.
    A grouped learning rate strategy is adopted. The backbone, gated fusion module, and attention pooling module share a base learning rate of $5 \times 10^{-5}$.   
    The $N_e$, $N_f$, and $\mu$ heads use learning rates of 0.25, 1.0, and 1.25 times the base rate, respectively.
    
    The total loss function is defined as a weighted combination of the losses from the three prediction heads. The loss for predicting the number of the singly excited states is given by a weighted cross-entropy loss with label smoothing $\epsilon = 0.10$ as
    \begin{equation}
    \mathcal{L}_{N_e} = \mathrm{CrossEntropy}(z_{n_e},\, y_{n_e}),
    \end{equation}
    where $z_{n_e}$ denotes the predicted logits and $y_{n_e}$ denotes the labels.
    The loss for selecting valid doubly excited states is defined using binary cross-entropy
    \begin{equation}
    \mathcal{L}_{N_f} = \mathrm{BCE}(\hat{k},\, k),
    \end{equation}
    where $\hat{k}$ denotes the predicted probabilities and $k$ denotes the labels. 
    The regression loss for the matrix of the transition dipole moment combines absolute error, relative error, and a correlation term, thereby constraining numerical accuracy, relative scaling, and global structural consistency
    \begin{equation}
    \mathcal{L}_{\mu} = \mathcal{L}_{\mu}^{\rm rel} + \lambda_{\rm abs}\,\mathcal{L}_{\mu}^{\rm abs} + \mathcal{L}_{\mu}^{\rm corr}.
    \end{equation}
    The absolute error term $\mathcal{L}_{\mu}^{\rm abs}$ emphasizes overall scale consistency and stabilizes multi-task training. After epoch 60, $\lambda_{\rm abs}$ is increased from $1.0$ to $5.0$ to place greater emphasis on the regression task.
    
    The total loss is defined as
    \begin{equation}
    \mathcal{L}_{\mathrm{tot}} = \lambda_{N_e}\,\mathcal{L}_{N_e} + \mathcal{L}_{N_f} + \mathcal{L}_{\mu}.
    \end{equation}
    {  During the first 60 epochs, the weight $\lambda_{N_e}$ is set to $1.0$. Since the $N_e$ prediction task is relatively simple, regularization is introduced to mitigate overfitting. Specifically, the $N_e$ head incorporates two dropout layers with rates $0.1$ and $0.2$, while the $N_f$ head employs two dropout layers, each with a dropout rate of $0.1$.
    Due to label smoothing, $\mathcal{L}_{N_e}$ remains non-zero even when the model achieves high prediction accuracy. To avoid overemphasizing this term, $\lambda_{N_e}$ is reduced to $0.5$ after epoch 60.}

\section{Training Results and Discussion}
\label{sec:Training}

\begin{figure}[h]
    \includegraphics[width=8.45cm]{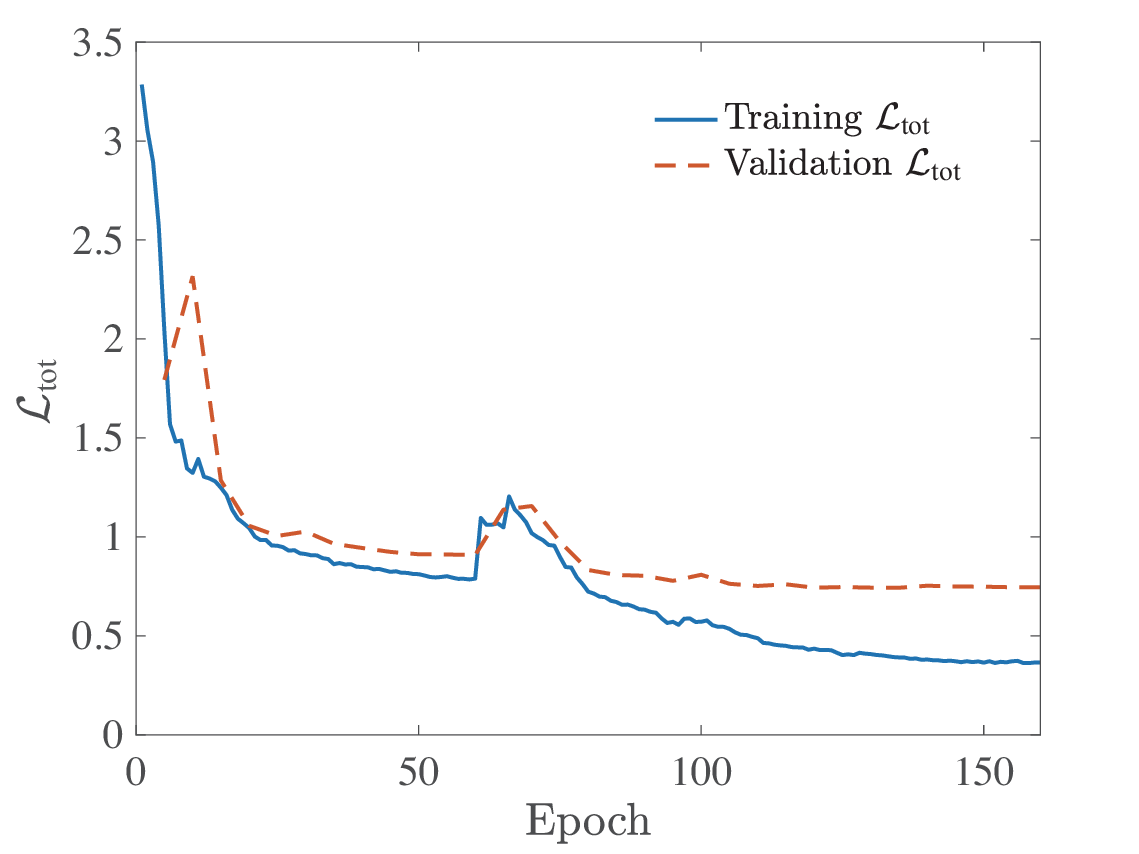}
    \centering
    \caption{The total loss $\mathcal{L}_{\mathrm{tot}}$ as a function of training epochs. The blue solid line denotes the training loss, while the orange dashed line denotes the validation loss.}
    \label{fig:TotalLoss}
\end{figure}

The dataset consists of 8000 $\vec{P}_{ZZZZ}$ {  2D IR spectra}, split into training (80\%) and validation (20\%) sets, with stratification by sub-dataset. Spec2DipoleNet is trained for 160 epochs on Google Colab using an NVIDIA A100-SXM4-80GB GPU, with validation performed every 5 epochs. 

The total loss $\mathcal{L}_{\mathrm{tot}}$ during training and validation is shown in Fig.~\ref{fig:TotalLoss}.
$\mathcal{L}_{\mathrm{tot}}$ decreases from 3.28 at initialization to 1.79 at the first validation point at epoch 5.
During the first 20 epochs, $\mathcal{L}_{\mathrm{tot}}$ exhibits a rapid decay and stabilizes around epoch 30. The validation loss follows a similar trend, although noticeable fluctuations appear in the early stage, including a transient increase despite continued reduction of the training loss. This behavior is likely due to the stochastic nature of early training and the application of data augmentation.
After epoch 60, the loss weights are adjusted and the learning rate enters the second annealing cycle. The training loss initially increases due to the enhanced contribution of $\mathcal{L}_{\mu}^{\rm abs}$, and subsequently decreases and stabilizes around $\mathcal{L}_{\mathrm{tot}} \approx 0.36$. The validation loss shows a similar but weaker decay, eventually stabilizing around $\mathcal{L}_{\mathrm{tot}} \approx 0.75$.

\begin{figure}[h]
    \includegraphics[width=8.45cm]{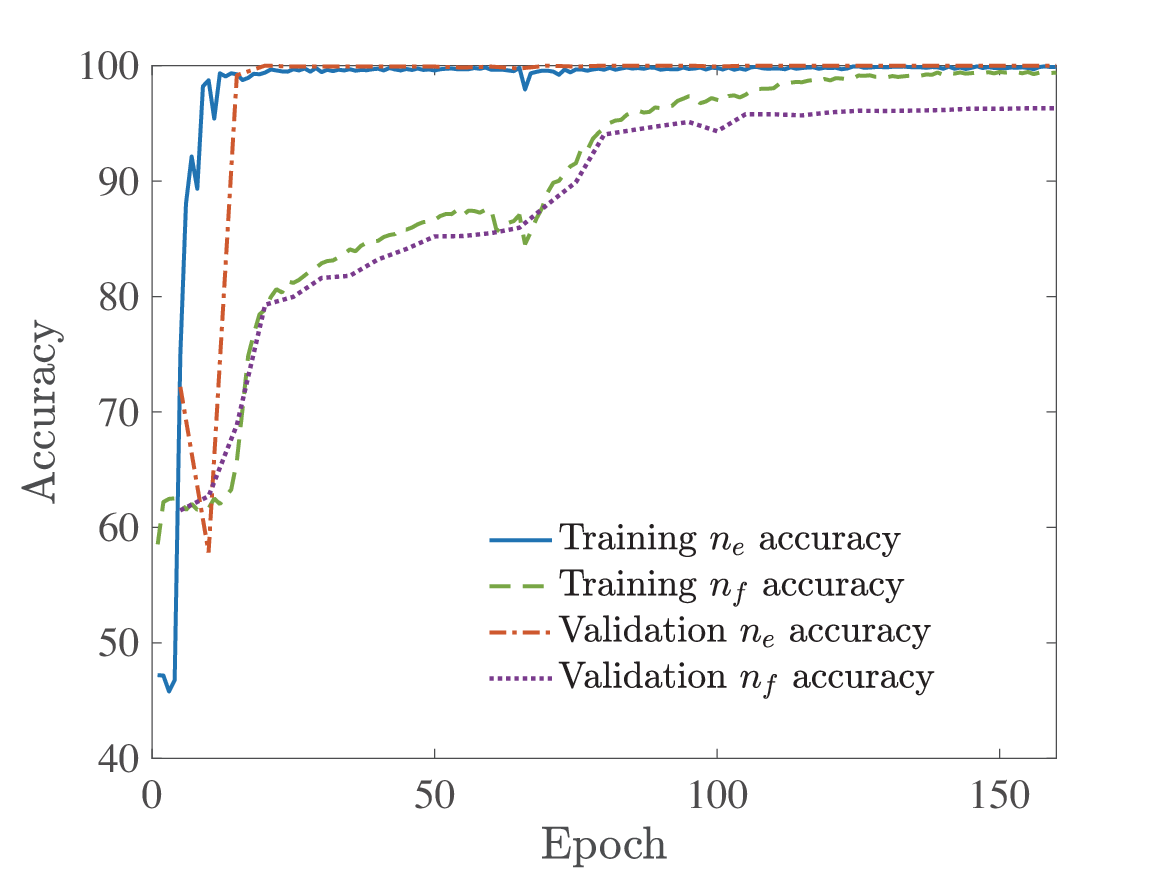}
    \centering
    \caption{Accuracy of the prediction for the number of the singly excited states $n_e$ and the presence of each doubly excited state on the training and validation set as a function of training progress. The blue solid and orange dashed lines denote the training and validation accuracy for $n_e$, respectively, while the green solid and purple dashed lines denote those for $n_f$.}
    \label{fig:AccuracyNef}
\end{figure}

Figure~\ref{fig:AccuracyNef} shows the training and validation accuracy of the $N_e$ and $N_f$ classification heads. The accuracy of $N_e$ increases rapidly in the early stages, reaching $99.33\%$ by epoch 12. Although the training accuracy initially exceeds the validation accuracy, both converge to $\sim 99\%$ after epoch 15 and remain stable thereafter.
In contrast, the $N_f$ task converges more slowly. This task involves binary classification over multiple doubly excited states, whose signatures in the {  2D IR spectra} are comparatively weaker, making it intrinsically more challenging than $N_e$. The accuracy of $N_f$ increases rapidly during the first 20 epochs, followed by a slower improvement until epoch 100. Beyond this point, the training accuracy continues to approach $100\%$, whereas the validation accuracy saturates at $96.30\%$.

After the loss weights are adjusted, the model places greater emphasis on minimizing the regression loss. Following a brief fluctuation over several epochs, the $N_f$ accuracy recovers and continues to improve, likely facilitated by the increased learning rate at the beginning of a new annealing cycle.

\begin{figure}[h]
    \includegraphics[width=8.45cm]{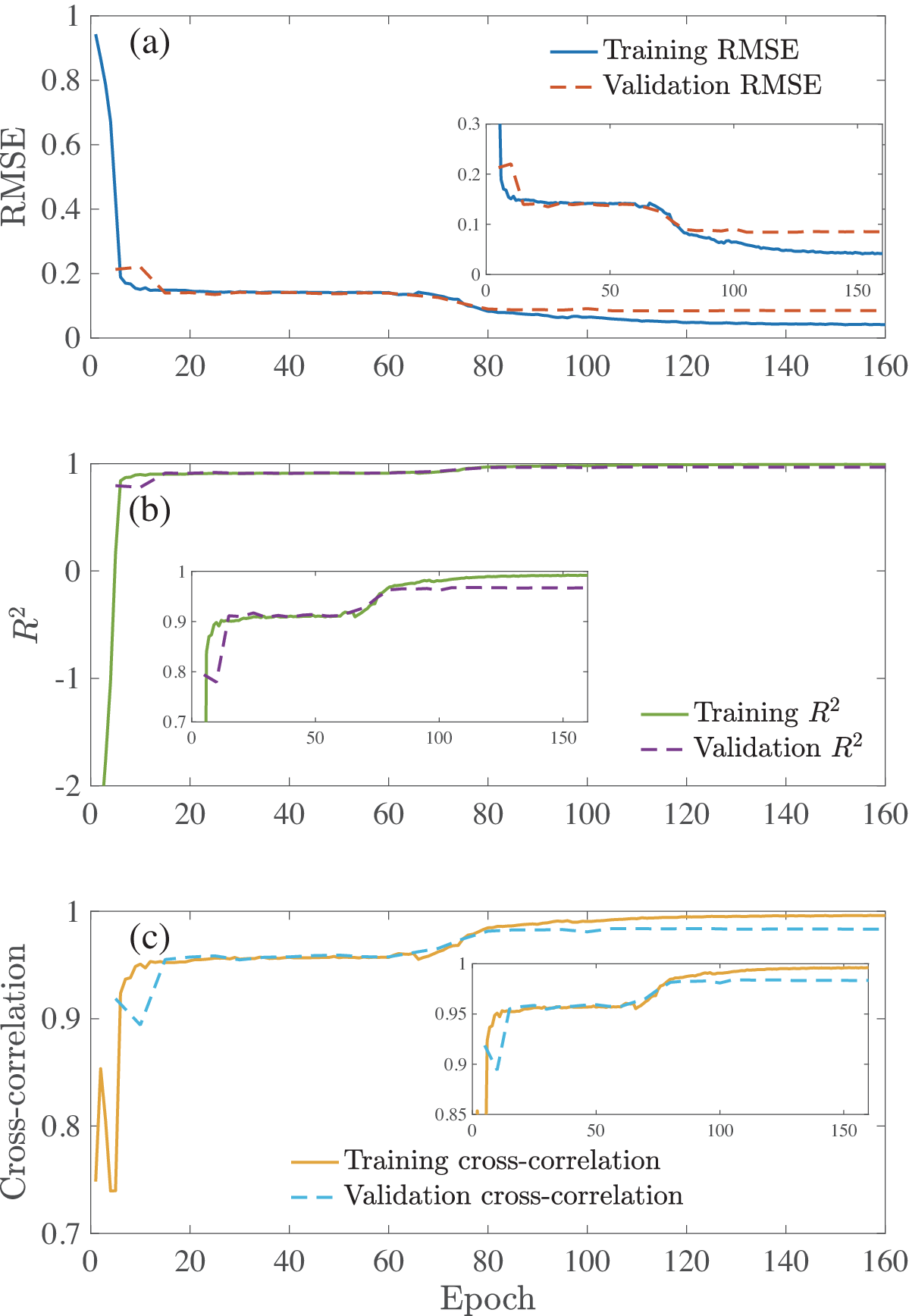}
    \centering
    \caption{RMSE, $R^2$, and NCC of the transition dipole matrix during training and validation. (a) RMSE for the training (blue solid) and validation (orange dashed) sets. (b) $R^2$ for the training (green solid) and validation (purple dashed) sets. (c) NCC for the training (yellow solid) and validation (blue dashed) sets.}
    \label{fig:muWhole}
\end{figure}

The performance of the $\mu$ regression head is evaluated using multiple {  dimensionless metrics}.
 Figure~\ref{fig:muWhole} presents the root-mean-square error (RMSE), the coefficient of determination $R^2$, and the normalized cross-correlation (NCC) for the predicted matrices of the transition dipole. These metrics quantify the error magnitude, the goodness of fit, and the structural similarity between predictions and ground truth, respectively. The definitions are given by
\begin{align}
\mathrm{RMSE} &=
\sqrt{\frac{1}{N}\sum_{i=1}^{N}({\mu}_i-\hat{{\mu}}_i)^2},\\
R^2 &=
1-\frac{\sum_{i=1}^{N}({\mu}_i-\hat{{\mu}}_i)^2}
{\sum_{i=1}^{N}({\mu}_i-\bar{{\mu}_i})^2},\\
\mathrm{NCC}(\hat{\mu}_i,\mu_i)
&=
\frac{(\hat{\mu}_i-\overline{\hat{\mu}}_i)\cdot(\mu_i-\overline{\mu}_i)}
{\|\hat{\mu}_i-\overline{\hat{\mu}}_i\|\,
\|\mu_i-\overline{\mu}_i\|}
\end{align}
where $\mu_i$ and $\hat{\mu}_i$ denote the ground-truth and the predicted matrices of the transition dipole for the $i$th sample, respectively, $\overline{\mu}_i$ and $\overline{\hat{\mu}}_i$ denote the element-wise mean matrices over all samples, and $N$ is the number of samples.
To compute these metrics, a mask is constructed based on the ground-truth values of $n_e$ and the valid doubly excited states, excluding the redundant matrix elements with zero-valued labels. In principle, the mask could be generated from the predicted quantities. However, this would introduce mismatches between different physical transitions in the metric evaluation. Using the ground-truth mask instead ensures a consistent comparison and provides a more reliable assessment of the regression performance.
In Fig.~\ref{fig:muWhole}(a)-(c), the RMSE, $R^2$, and the NCC for the matrix of the transition dipole significantly improve during the first 10 epochs. The RMSE decreases from 0.94 at epoch 1 to 0.15 at epoch 10. $R^2$ increases from -2.98 at epoch 1 to 0.90 at epoch 12. The NCC increases from 0.75 at epoch 1 to 0.95 at epoch 10.
These global metrics are improved after epoch 60, when the weight of the loss function is adjusted, and further improved around epoch 85. Before epoch 85, the values and trends of these metrics on both sets closely matched. After epoch 85, while the model's accuracy continues to improve on the training set, the validation set remains stable.
The insets of Fig.~\ref{fig:muWhole} provides a detailed view of the local metrics. After epoch 15, the RMSE on both the training and validation set saturates at approximately $0.14$, while $R^2$ stabilizes around $0.91$ and the NCC around $0.96$. All metrics show consistent improvement after epoch 60, coinciding with the reweighting of the loss terms, and converge around epoch 85.
Beyond epoch 85, training performance continues to improve, whereas validation metrics remain stable. The RMSE, $R^2$ and the NCC on the validation set finally reach $0.085$,  $0.97$ and $0.98$, respectively. 

\begin{figure*}[htbp]
    \includegraphics[width=17.4cm]{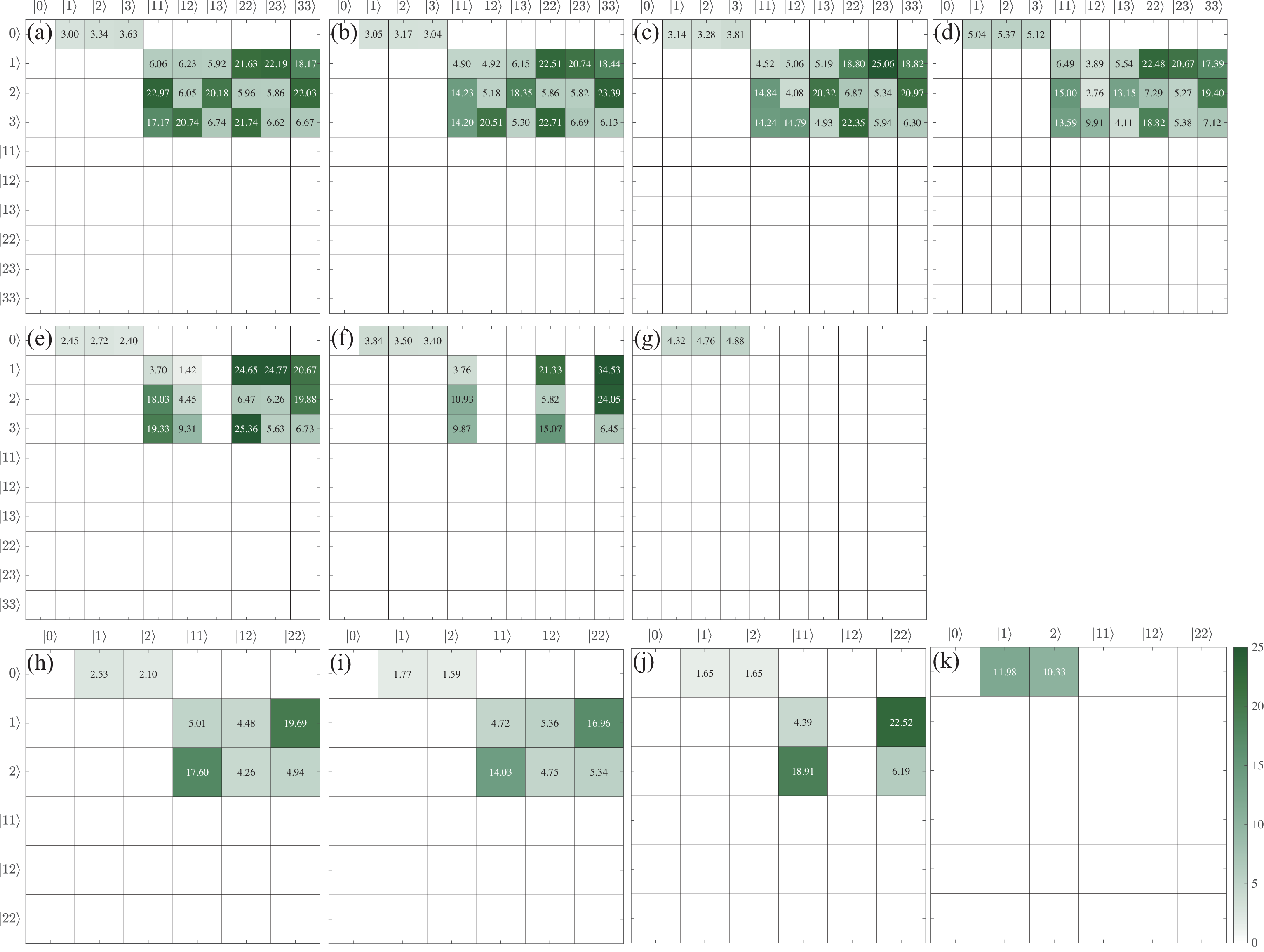}
    \centering
    \caption{Relative error matrices of the transition dipole operator. (a)--(d) correspond to systems with 10 to 7 energy levels. (e)--(g) correspond to systems with 6 to 4 energy levels, and (h)--(k) correspond to systems with 6 to 3 energy levels. 
    The horizontal and vertical axes label the ground state, singly excited states ($n_e=3$ or $2$), and doubly excited states ($n_f=6,5,4,3$, $2,1,0$, or $3,2,1,0$ for the respective panels). The average relative error for each matrix element is indicated within the corresponding color block.}
    \label{fig:RelALL}
\end{figure*}

Figure~\ref{fig:RelALL} presents the element-wise relative errors of the predicted matrices of the transition dipole for systems with different numbers of energy levels. The relative error is defined as
\begin{equation}
\label{relError}
\mathrm{RelError} = \frac{\hat{\mu}_i - \mu_i}{|\mu_i| + \epsilon},
\end{equation}
with $\epsilon = 0.001$.
In computing these statistics, entries with mismatched predicted and ground-truth masks are excluded, since such discrepancies arise from incorrect predictions of the level structure rather than errors in the associated matrix elements.

%For systems with $n_e=3$ singly excited states and up to six doubly excited states, {cf. Fig.~\ref{fig:RelALL}(a)-(d)}, the transitions from the ground to singly excited states exhibit the lowest relative errors $\sim3\%$, followed by the allowed transitions from the singly  to doubly excited states $\sim5\%$--$7\%$, while the forbidden transitions show significantly larger errors $\sim17\%$--$23\%$. 

For systems with $n_e=3$ singly excited states and up to six doubly excited states, {cf. Fig.~\ref{fig:RelALL}(a)-(d)}, the transitions from the ground to singly excited states exhibit the lowest relative errors of $\sim 3\%$. {  Importantly, the allowed transitions between the singly and doubly excited manifolds, which contribute to excited-state absorption features, are also recovered with relatively low errors of $\sim 5\%$--$7\%$. The substantially larger errors of $\sim 17\%$--$23\%$ are mainly associated with the nominally forbidden, or weakly allowed, transitions. Thus, the reduced accuracy is not uniform across all excited-state transitions.}

The larger errors for the weakly allowed transitions arise partly from the small magnitudes of their transition dipole matrix elements, which produce weaker spectral features. 
Moreover, since these elements appear in the denominator of Eq.~(\ref{relError}), their small magnitudes amplify the relative error.
{  In other words, even a small absolute deviation in the predicted matrix element can result in a large relative error when the true transition dipole strength is very small.}
In addition, partial spectral overlap between different transitions further limits the prediction accuracy.
As shown in Fig.~\ref{fig:ZZZZ2DS}, distinct transitions may share identical or similar frequencies, resulting in overlapping peaks in the {  2D IR spectra}. This overlap hinders the disentanglement of individual contributions and imposes an intrinsic limitation on the regression accuracy.

As the total number of energy levels decreases, similar trends persist, while additional variability arises from the smaller sample sizes, the energy-level degeneracy, and the finite pulse bandwidth. 
For the systems with fewer doubly excited states, {cf. Fig.~\ref{fig:RelALL}(e)-(g)}, the relative errors of the transitions from the ground to singly excited states remain below $5\%$, consistent with the above trends. Larger deviations, such as the $34.53\%$ error, arise in the case of the forbidden transitions, where the corresponding transition is underrepresented in the dataset because of the limited number of samples
For systems with $n_e=2$, {cf. Fig.~\ref{fig:RelALL}(h)-(k)}, the simpler energy-level structure leads to lower errors for the transitions from the ground to singly excited states, lower by approximately $1$--$2\%$ compared to the case with $n_e=3$. 
The transitions from the singly  to doubly excited states follow the same stratification observed in {Fig.~\ref{fig:RelALL}(a)-(d)}, where allowed transitions exhibit relatively low errors $\sim4\%$--$5\%$, while the forbidden transitions remain higher $\sim14\%$--$23\%$. However, in the simplest configurations, the relative errors increase to $\sim10\%$--$12\%$. This is likely due to the reduced spectral complexity, where fewer distinguishable features are available to constrain the regression, making the prediction more sensitive to the noise and normalization.

\section{Conclusion}
\label{sec:Conclusion}

We have developed Spec2DipoleNet, a physics-informed multi-task network that infers transition dipole matrices directly from $\vec{P}_{\mathrm{ZZZZ}}$ {  2D IR spectra}. Trained on 8000 simulated spectra from systems containing $N=3$--$10$ energy levels, the model achieves high accuracy in both level-structure classification and transition dipole regression. The validation accuracy reaches $\sim 99\%$ for $n_e$ and $96.30\%$ for $n_f$. For transition dipole regression, the validation RMSE, $R^2$, and NCC saturate at approximately $0.085$, $0.97$, and $0.98$, respectively.

In the evaluation, errors from the prediction for the structure and the regression for the transition dipole are separated, allowing the intrinsic performance of the regression task to be assessed independently. The results reveal a clear physically-driven error stratification. In the ten-level system, the transitions from the ground to the singly excited states achieve relative errors of $3.00\%$--$3.63\%$, and the allowed transitions from the singly  to doubly excited states remain at $5.86\%$--$6.74\%$, while the forbidden transitions increase to $17.17\%$--$22.97\%$. Furthermore, simpler spectra do not necessarily improve regression accuracy. For $n_e=2$, the relative errors in the simplest $n_f=0$ case rise to $11.98\%$ and $10.33\%$, indicating a fundamental limitation set by the spectral identifiability. These findings demonstrate that the recovery of the transition dipole from the {  2D IR spectra} is governed not only by the model capacity, but also by the physical identifiability of the underlying inverse problem.

Overall, this work demonstrates that deep learning, when combined with physically-structured modeling and careful evaluation, provides a powerful approach for solving inverse problems in multidimensional spectroscopy. The framework can be naturally extended to more complex settings, including multiple waiting times and realistic experimental conditions, offering a promising direction for spectral analysis.

\begin{acknowledgments}

This work is supported by the National Natural Science Foundation of China under Grant no. 62461
160263, Innovation Program for Quantum Science and Technology under Grant no. 2023ZD0300200,
and Guangdong Provincial Quantum Science Strategic Initiative under Grant no. GDZX2505004.

\end{acknowledgments}

\bibliography{ref_main.bib}

\end{document}